\documentclass[osajnl,twocolumn,showpacs,superscriptaddress,10pt]{revtex4-1} 
\usepackage{amsmath,amssymb,graphicx}
\begin{document}

\title{Design of a hybrid silicon-plasmonic co-propagant integrated coherent perfect absorber}

\author{Simone Zanotto}
\affiliation{Dipartimento di Elettronica, Informazione e Bioingegneria, Politecnico di Milano, Piazza Leonardo da Vinci 32, 20133 Milano, Italy}
\affiliation{Present address: CNR-INO, Via Nello Carrara 1, 50019 Sesto Fiorentino (FI), Italy}
\author{Andrea Melloni}

\affiliation{Dipartimento di Elettronica, Informazione e Bioingegneria, Politecnico di Milano, Piazza Leonardo da Vinci 32, 20133 Milano, Italy}

\begin{abstract}By a hybrid integration of plasmonic and dielectric waveguide concepts, it is shown that coherent perfect absorption can be achieved in a co-propagant coupler geometry. The device holds promises for classical and quantum signal processing. First, the operating principle of the proposed device is detailed in the context of a more general 2x2 lossy coupler formalism. Then, it is shown how to tune the device in a wide region of possible working points, its broadband operation, and the tolerance to fabrication uncertainties. Finally, a complete picture of the electromagnetic modes inside the hybrid structure is analyzed, shining light onto the potentials which the proposed device holds in view of other applications like nonlinear optics, polarization control, and sensing. 
\end{abstract}


\maketitle 

Losses occurring in optical systems are either strongly desired or carefully avoided, depending on the specific device under consideration. For instance, detectors and solar cells fall in the first category, whereas integrated optical elements such as waveguides and couplers usually belong to the second one. However, far from being an exhausted field of study, the analysis of absorption in linear optical systems recently attracted the community attention, mainly in connection with the non-trivial phenomenon known as coherent perfect absorption (CPA). In essence, CPA consists of the complete absorption of two coherent beams by an otherwise partially absorbing sample or device. As long as monochromatic stationary waves are concerned, CPA has the remarkable property of being the time-reversal counterpart of a laser at threshold \cite{ChongPRL2010}; even richer physical analogies with lasers beyond threshold occur when transient or chaotic optical fields are involved \cite{LonghiPRA2012}.

While most of the studies about CPA focused upon structures working with free-space optical radiation \cite{CaoScience2011, NohPRL2012, FengPRB2012, PuOE2012, YoonPRL2012, ZhangOE2014, RaoOL2014, SociNatComm2015}, there are only few reports which treat the phenomenon within the integrated optics framework \cite{GroteOL2013, BruckOE2013}, despite the integration of a CPA device on a scalable, on-chip platform could have a deep impact in different fields of optical science and technology. Indeed, CPA can be exploited both for coherent processing of classical signals and for quantum optical computing purposes \cite{DeLeonJSTQE2012, HeeresNatNano2013}. On the first side, CPA-based detection schemes for phase-keyed signals have been devised \cite{GroteCLEO2015}, while on the second side many opportunities related to the statistics of individual photons are still to be explored. Possible experiments generalizing those with cascades of non-lossy beam splitters \cite{TillmannNatPhot2013, SpagnoloNatPhot2014} could be envisaged, exploiting the apparent two-photon absorption in a linear medium which is enabled by CPA \cite{BarnettPRA1998}.

Here we propose, design and analyze a co-propagant CPA device based on the hybrid integration of a plasmonic element into a 2x2 coupler implemented on the silicon-on-insulator platform. Thanks to the co-propagant geometry, it can allow for a reduction of the global footprint with respect to the counter-propagant approach analyzed in Ref.~\cite{BruckOE2013}. Our device expands the current library of hybrid devices \cite{MojahediLPR2014}, sharing with them the potentials for biosensing, nonlinear optics \cite{KleinScience2006, CaiScience2011, SederbergPRL2015, SegalNatPhot2015}, and even for detection, if the hot carriers generated in the metal are collected through Shottky barriers \cite{KnightScience2010, BrongersmaNatNano2015, KonstantatosACSPhot2015}.

The concept of a generic co-propagant CPA coupler is schematized in Fig.~1 (a-b). 
\begin{figure}[htbp]
\centerline{\includegraphics[width=\columnwidth]{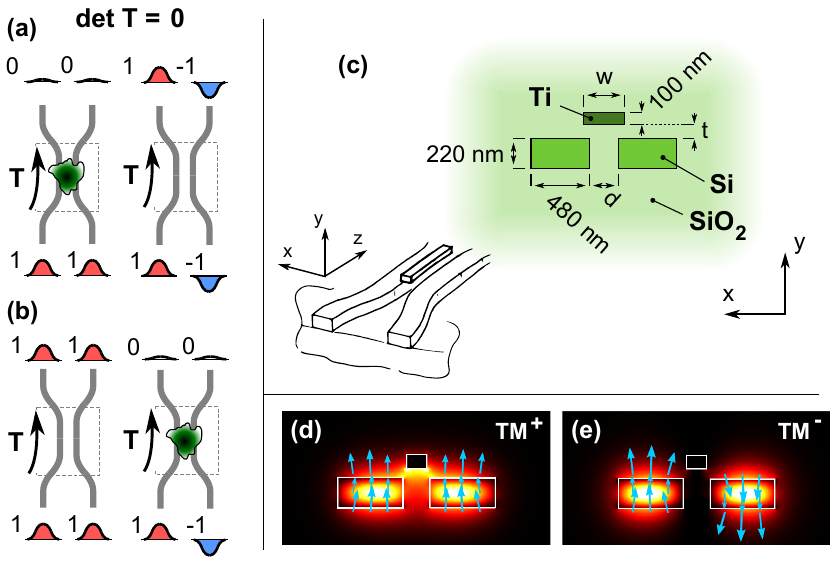}}
\caption{Panels (a-b): schematical operation of an axisymmetric co-propagant CPA coupler. With the symmetry constraint, CPA can either occur for in-phase or out-of-phase inputs. Panel (c): a CPA coupler implemented through hybrid integration of a metal component with silicon photonic coupled waveguides. Its operating principle is the difference between the field distribution among even and odd TM supermodes [panels (d-e)]. The large extinction of the TM$^+$ supermode leads to a situation such that of panel (a).}
\end{figure}
Considering structures with a symmetry axis, two distinct possibilities exist. Either the system fully absorbs the symmetric input state, or it absorbs the antisymmetric one. This is a consequence of the CPA condition, here written $\det T = 0$, being $T$ the $2 \times 2$ coupler transfer matrix. In other words, one of the $T$-matrix eigenvalue is zero, and its eigenstate can be either the symmetric or the antisymmetric one. In addition, a common requirement (referred to as \textit{coherent perfect transparency}, CPT) is that the second eigenvalue has unity modulus. This means that the input state orthogonal to the absorbed one is perfectly transmitted; this situation is also depicted in the Figure. Notice that at this stage the discussion is about the states of the \textit{coupler} in its whole, not of the \textit{coupling region}, and that there are in principle many routes to realize a CPA co-propagant coupler.  A simple method to synthesize a CPA coupler is to engineer the absorption of the symmetric and antisymmetric supermodes of the coupling region, by properly modifying an ordinary directional coupler. A geometry which implements this idea is depicted in Fig.~1 (c). Two standard $220 \times 480$ nm Silicon waveguides are buried into a SiO$_2$ cladding, and a Titanium stripe is placed on the axis symmetry of the coupler, vertically spaced by $t$ from the waveguide top plane. Titanium is chosen as a high-loss metal compatible with silicon photonics processing. The waveguides are separated by $d$, and the metal stripe width is $w$. Later on in the article, a thorough analysis of the effect of the parameters will be detailed.

Fig.~1 (d-e) depicts the computed symmetric and antisymmetric (quasi-) transverse-magnetic supermodes of the coupling region, labeled respectively TM$^+$ and TM$^-$. As it can be directly observed by the color map, which represents the $z$-component of the Poynting vector, the TM$^+$ mode shows some energy flux in close vicinity of the metal stripe. This eventually ends up in a large extinction coefficient for that mode. Instead, in the TM$^-$ supermode, which must fulfil field antisymmetry with respect to the vertical symmetry axis, there is not electromagnetic energy guided in close vicinity of the metal stripe, and hence a lower absorption is expected. 
The complex propagation constants of the coupler supermodes $\beta^{\pm}$ are the key quantities entering the lossy coupler $T$-matrix and the CPA/CPT conditions. The $\beta^{\pm}$ are defined in terms of the complex effective refractive indices of the coupler supermodes $n_{\mathrm{eff}}^{\pm}$ as $\beta^{\pm} = 2 \pi (\mathrm{Re}~n_{\mathrm{eff}}^{\pm} + i \mathrm{Im}~n_{\mathrm{eff}}^{\pm}) /\lambda_0 $, being $\lambda_0$ the vacuum wavelength. Neglecting any spurious effect (i.e., reflections,  radiation, and imperfect modal overlap) at the transition regions, one gets

\begin{equation}
 T = \begin{array}{cc}
 
  \left( \begin{array}{cc}
     t_{11} & t_{21} \\
     t_{12}  & t_{22} 
    \end{array} \right) ;
    &
      \begin{array}{c}
     t_{11} = t_{22} = \left( e^{i \beta^+ L} +  e^{i \beta^- L} \right) /2 \\
     t_{12} = t_{21} = \left( e^{i \beta^+ L} -  e^{i \beta^- L}  \right) /2
    \end{array} 
    
    \end{array}
\label{Tgen}
\end{equation}
where $L$ is the length of the coupling region. Simple algebra leads to $\det T = e^{i \beta^+ L} e^{i \beta^- L}$, and at a first sight a proper coherent \textit{perfect} absorption condition cannot be met. However, if either $\mathrm{Im}~\beta^+ L$ or $\mathrm{Im}~\beta^- L$ is large, the system is close to CPA. Better yet, if $\mathrm{Im}~\beta^+ \gg \mathrm{Im}~\beta^-$ (or the reverse), the system is both close to CPA and to CPT. Indeed, up to a global phase, in this limit its transmission matrix reads
\[
 T = \left( \begin{array}{cc}
     1/2  & \pm 1/2\\
     \pm 1/2 &  1/2 
    \end{array} \right)
\]
where the sign is determined by the relative magnitude of $\mathrm{Im}~\beta^{\pm}$. Referring to the specific device geometry analyzed in the present work, $\mathrm{Im}~\beta^+ \gg \mathrm{Im}~\beta^-$, and the minus sign applies. 
Notice that, when a single arm of the coupler is excited, the output signals are either in phase or out of phase by $\pi$, as opposed to ordinary lossless directional couplers where a $\pi/4$ dephasing is observed. 

To quantify how an engineered lossy coupler is close to CPA and CPT it is useful to rely on two quantities, extinction ratio (ER) and insertion loss (IL). In this framework they are defined as $\mathrm{IL} = -10 \log_{10} (I_{\mathrm{out; max}}/I_{\mathrm{in}})$ and $\mathrm{ER} = -10 \log_{10} (I_{\mathrm{out; min}}/I_{\mathrm{out; max}})$, where $I_{\mathrm{out; min, max}}$ are the total minimum and maximum intensities available at the output arms of the coupler. Minimum and maximum can be reached by sweeping the relative phase of the inputs; notice that in the 2x2 axisymmetric configuration these extrema are reached when signals of equal intensity $I_{\mathrm{in}}/2$ are employed as inputs (balanced inputs; the behaviour of CPA devices without axial symmetry are discussed in \cite{BaldacciOE2015}). 

Insertion loss and extinction ratio can be interpreted both from an \textit{internal} and an \textit{external} point of view. In the first case, the focus is on the energy available inside the coupler, which can be harnessed, for instance, for heating, for detection, or for mechanical motion \cite{taylor2015enhanced}. In the second case, the main point is about the energy which is delivered at the output ports, which can be subsequently detected or rerouted into a forthcoming circuit. In both cases, IL should be as small as possible, while ER should be maximized. Clearly, IL and ER are connected to the T-matrix elements: when the generic form for the T-matrix is considered, one gets $I_{\mathrm{out; max}} = 2|t_{11}\pm t_{12}|^2$, where the sign is determined by whether the coupler has a symmetric or antisymmetric CPA \footnote{Note that symmetry and reciprocity imply $t_{11} = t_{22}$ and $t_{12} = t_{21}$.}. When instead the co-propagant CPA coupler based on differential losses between supermodes is analyzed, in the case $\mathrm{Im}~\beta^+ > \mathrm{Im}~\beta^-$ the expressions $\mathrm{IL} = 20 \log_{10} \exp \left( \mathrm{Im}(\beta^-) L\right) $ and $\mathrm{ER} = 20 \log_{10} \exp \left( \mathrm{Im}(\beta^+ - \beta^-) L \right)$ are obtained. Here, while IL and ER linearly scale with length, their ratio do not, and can be considered as a figure of merit for the CPA coupler.

As the geometrical parameters and the operating wavelength are varied, the device of Fig.~1(c) shows different values of IL and ER. For instance, if $t = 70\ \mathrm{nm}$, $w = 150\ \mathrm{nm}$, $d = 200\ \mathrm{nm}$, and $\lambda = 1.55\ \mu \mathrm{m}$ one obtains $\mathrm{IL} = 0.11\ \mathrm{dB}/\mu \mathrm{m}$ and $\mathrm{ER} = 1.23\ \mathrm{dB}/\mu \mathrm{m}$. Hence, devices with lengths of the order of tens of microns can implement a satisfactory extinction factor. 

When the parameters $w$, $t$ and $d$ are varied, the results reported in Fig.~2 are obtained: in the left column, $t$ and $w$ are varied while $d = 200\ \mathrm{nm}$ is kept fixed; in the right column, instead, $d$ and $w$ are swept while $t = 70\ \mathrm{nm}$ is kept constant. The operating wavelength is fixed at $1.55\ \mu \mathrm{m}$. In each column, the dashed lines highlight the value of the parameter which is kept constant in the other column; the black dots, instead, represents the parameters which will be employed for Fig.~3. 
\begin{figure}[htbp]
\centerline{\includegraphics[width=\columnwidth]{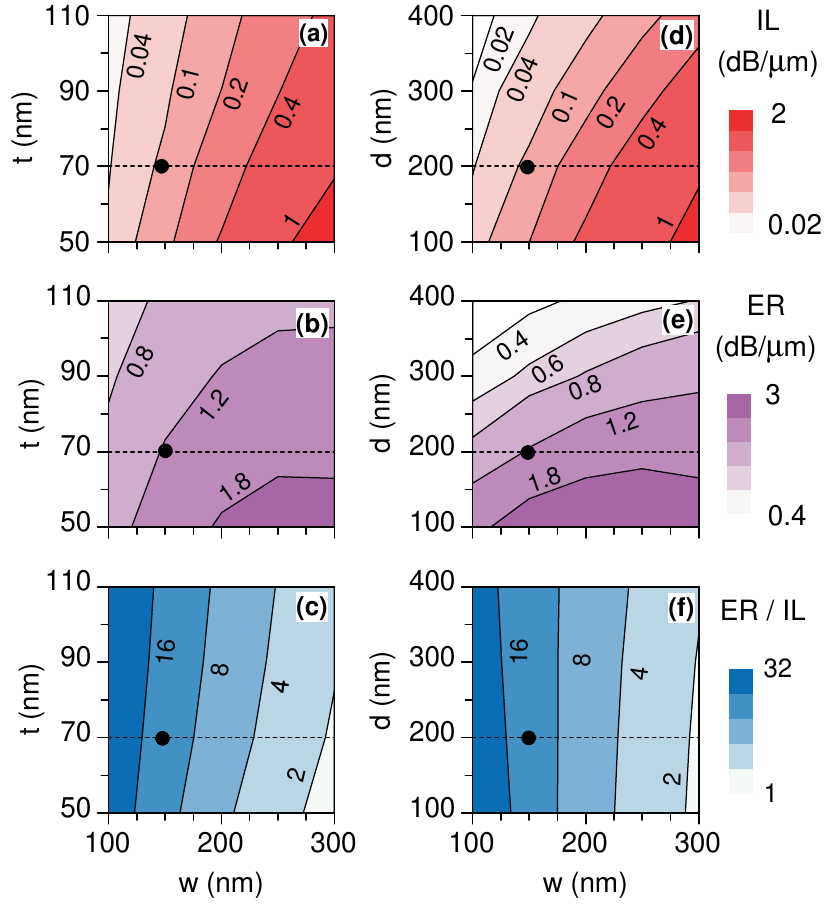}}
\caption{Dependence of the CPA coupler key parameters insertion loss and extinction ratio (IL ad ER) with respect to the geometrical parameters $t$, $w$, and $d$. In the left panels $d = 200\ \mathrm{nm}$, while in the right ones $t = 70\ \mathrm{nm}$. IL and ER can be tailored by properly tuning $t$, $w$, and $d$.}
\end{figure}
The rows of Fig.~2 group insertion loss, extinction ratio, and the ``figure of merit'' $\mathrm{ER}/\mathrm{IL}$. Analyzing Fig.~2 three main features stand out. First, insertion loss is strongly affected by $w$ and weakly by $t$ and $d$. Values as small as $0.02\ \mathrm{dB}/\mu \mathrm{m}$ can be obtained for $w \simeq 100\ \mathrm{nm}$, $d \simeq 400\ \mathrm{nm}$. Second, the effect on the extinction ratio is mostly ascribed to the waveguide spacing $d$, which, when brought down to $d \simeq 100\ \mathrm{nm}$, leads to ER's almost as large as $3\ \mathrm{dB}/\mu \mathrm{m}$. Third, the combined effect of the above lead to an overall insensitivity of the ratio $\mathrm{ER}/\mathrm{IL}$ upon variations of both $t$ and $d$, whereas it is largely affected by $w$. These features can be exploited to fit the CPA coupler to the desired working point, as well as to cope with the tolerance issues inherent in the fabrication processes. 

It will now be shown that the device is also robust with respect to wavelength and size perturbations. The data reported in Fig.~3 (a-b) are obtained for fixed values of the physical dimensions (reported in the inset), while the wavelength is swept within a wide band relevant for data- and telecommunication purposes. The insertion loss increases with the wavelength, from about $0.06\ \mathrm{dB}/\mu \mathrm{m}$ at $1350\ \mathrm{nm}$ to $0.13\ \mathrm{dB}/\mu \mathrm{m}$ at $1650\ \mathrm{nm}$. Meanwhile, the extinction ratio increases from $0.45\ \mathrm{dB}/\mu \mathrm{m}$ at $1350\ \mathrm{nm}$ to $1.4\ \mathrm{dB}/\mu \mathrm{m}$ at $1650\ \mathrm{nm}$: in the whole band under analysis, $\mathrm{ER}$ is almost ten times larger (in dB scale) with respect to $\mathrm{IL}$ (see Fig.~3 (a)). It should be noticed that the (adimensional) ratio $\mathrm{ER}/\mathrm{IL}$ does not depend on the physical length $L$ of the coupler, which can hence be tuned to match the specific request of insertion loss or extinction ratio, without affecting the overall performance of the device described instead by $\mathrm{ER}/\mathrm{IL}$. 

When the misalignment $\delta$ of the metal stripe is instead considered as the tolerance parameter, the data in Fig.~3 (c-d) are obtained. Here, the wavelength is fixed at $1550\ \mathrm{nm}$, and the component sizes are the same as above. 
\begin{figure}[htbp]
\centerline{\includegraphics[width=\columnwidth]{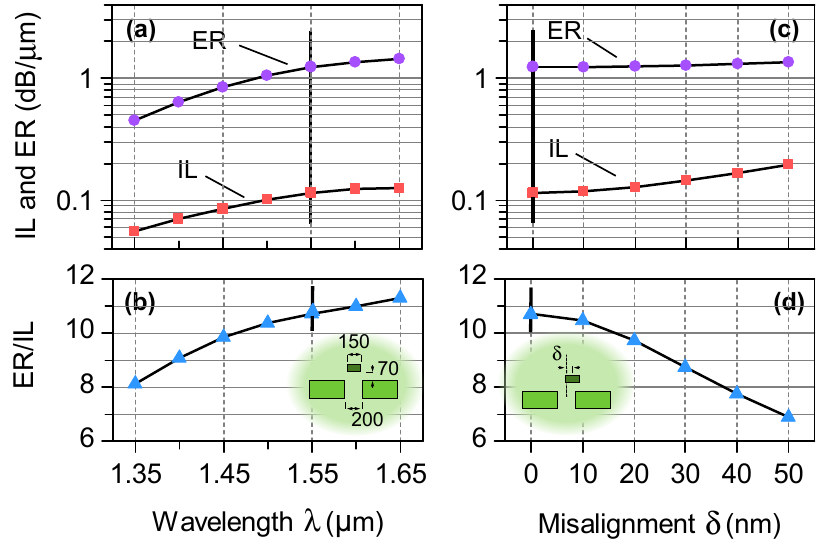}}
\caption{Panels (a-b): insertion loss (IL) and extinction ratio (ER) of a hybrid coupler, as a function of the wavelength, for the geometrical parameters reported in the inset. Large ER and small IL means that the device is close to CPA and CPT (see text). IL and ER are not strongly affected by the wavelength. Even smaller is the effect on the ratio ER/IL. Similar observations follow from the analysis of the tolerance with respect to the metal stripe misalignment, panels (c-d).}
\end{figure}
As it can be expected, the performance metric $\mathrm{ER}/\mathrm{IL}$ decreases when $\delta$ increases; however, even for the quite large value $\delta = 50\ \mathrm{nm}$, $\mathrm{ER}/\mathrm{IL}$ is as large as 7 units. The main reason is the increase of $\mathrm{IL}$, while $\mathrm{ER}$ is essentialy unaffected by the misalignment. While the tolerance of the CPA coupler with respect to the wavelength could have been expected, being the device not based on a resonance mechanism, the robustness with respect to the misalignment is more surprising. Indeed, symmetry is the operating principle of the device, and its breaking could have been suspected of a significant degradation of the performance.   

The final section of this article is devoted to a full modal analysis of the hybrid coupler, which gives further insights into the operating potentials of the proposed structure, possibly in view of applications not strictly related to the coherent perfect absorption mechanism. Fig.~4 represents a modal map of the hybrid coupler: the real part of the supermodes effective refractive index is plotted on the $x$-axis, while the imaginary part is plotted on the $y$-axis. 
\begin{figure}[htbp]
\centerline{\includegraphics[width=\columnwidth]{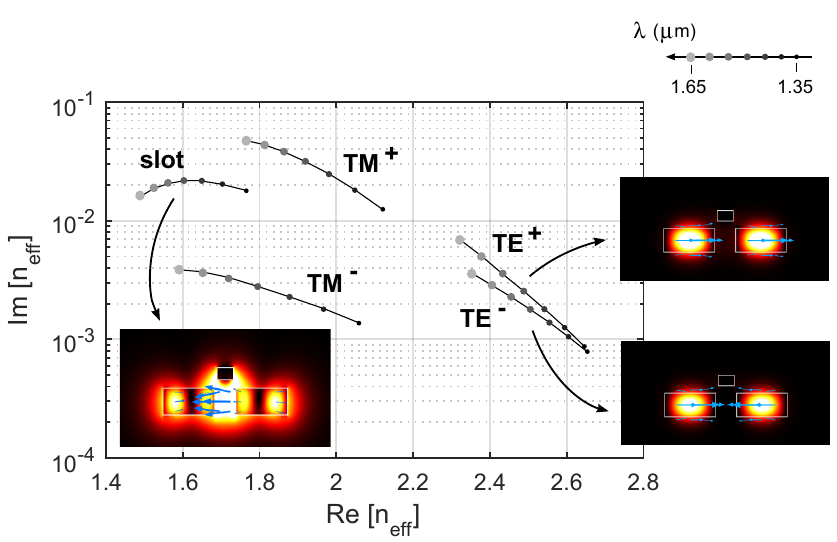}}
\caption{Modal map in the complex effective refractive index plane for the hybrid silicon-plasmonic coupler. TE and TM modes appear as pairs of supermodes with opposite symmetries, while a slot-like mode (hybridized with a plasmonic component) exists as a singlet. All the modes exist in the $(1.35, 1.65)\ \mu \mathrm{m}$ wavelength range, without mixing or crossing effects. In the insets, the color map represents the out-of-plane component of the Poynting vector, while the blue arrows represent the modal transverse electric field. The strong difference between $\mathrm{Im}~n_{\mathrm{eff}}$ for the TM modes is at the hearth of the CPA operation.}
\end{figure}
Parameters $w = 150\ \mathrm{nm}$, $t = 70\ \mathrm{nm}$, and $d = 200\ \mathrm{nm}$ have been chosen, while the wavelength is swept from $1.35$ to $1.65\ \mu \mathrm{m}$ (i.e., the same as in Fig.~3 (a-b)). The modes are grouped according to their symmetry and polarization: the field profiles of TM$^{\pm}$ modes is that already represented in Fig.~1 (d-e), while that of the other modes are reported here as picture insets. This representation immediately reveals several features of the studied structure. First, within the considered wavelength range, the complex effective index of each mode evolve continuously, indicating that no mode crossing and mixing occur. Second, the TM$^{\pm}$ modes exhibit a large splitting between $\mathrm{Im}~n_{\mathrm{eff}}$. This is a graphical way of understanding the co-propagating CPA effect, since, as it was highlighted above, a condition which implies (quasi-) CPA is that $\mathrm{Im}~\beta^{+} \gg  \mathrm{Im}~\beta^{-}$. Notice that, instead, TE$^{\pm}$ modes have a small splitting between $\mathrm{Im}~n_{\mathrm{eff}}$: in essence, the hybrid coupler acts as a \textit{polarization-sensitive CPA device}. Third, in addition to TE and TM modes, the device exhibits a peculiar dielectric slot - surface plasmon hybrid mode, which exists in a single symmetry configuration and which cannot be thereby exploited in view of CPA. However, thanks to the strong field confinement in the cladding region and in the vicinity of the metal stripe, this mode could be harnessed for switching purposes in nonlinear materials, opening up the perspective for further functionalities.

In conclusion, the design of an integrated optical component which implements coherent perfect absorption and transparency (CPA and CPT) is presented. The layout exploits hybridization of dielectric and plasmonic modes in a co-propagant coupler realized on the silicon photonic platform. By harnessing the tight field confinement together with the metal losses, a large extinction ratio and small insertion losses can be devised. A numeric analysis revealed that different working point can be properly set by tuning certain structure parameters, and that the structure is tolerant to fabrication uncertainties. The co-propagant CPA/CPT device, whose general behaviour in terms of 2x2 transfer matrices is also discussed, can be included in complex networks to elaborate classical and optical signals, offering new functionalities like signal demodulation and apparent single-photon non linearities in a single, small-footprint building block.


\begin{thebibliography}{10}
\newcommand{\enquote}[1]{``#1''}

\bibitem{ChongPRL2010}
Y.~D. Chong, L.~Ge, H.~Cao, and A.~D. Stone, \enquote{Coherent perfect
  absorbers: Time-reversed lasers,} Phys. Rev. Lett. \textbf{105}, 053901
  (2010).

\bibitem{LonghiPRA2012}
S.~Longhi and G.~Della~Valle, \enquote{Coherent perfect absorbers for
  transient, periodic, or chaotic optical fields: Time-reversed lasers beyond
  threshold,} Phys. Rev. A \textbf{85}, 053838 (2012).

\bibitem{CaoScience2011}
W.~Wan, Y.~Chong, L.~Ge, H.~Noh, A.~D. Stone, and H.~Cao,
  \enquote{Time-reversed lasing and interferometric control of absorption,}
  Science \textbf{331}, 889--892 (2011).

\bibitem{NohPRL2012}
H.~Noh, Y.~Chong, A.~D. Stone, and H.~Cao, \enquote{Perfect coupling of light
  to surface plasmons by coherent absorption,} Phys. Rev. Lett. \textbf{108},
  186805 (2012).

\bibitem{FengPRB2012}
S.~Feng and K.~Halterman, \enquote{Coherent perfect absorption in
  epsilon-near-zero metamaterials,} Phys. Rev. B \textbf{86}, 165103 (2012).

\bibitem{PuOE2012}
M.~Pu, Q.~Feng, M.~Wang, C.~Hu, C.~Huang, X.~Ma, Z.~Zhao, C.~Wang, and X.~Luo,
  \enquote{Ultrathin broadband nearly perfect absorber with symmetrical
  coherent illumination,} Opt. Express \textbf{20}, 2246--2254 (2012).

\bibitem{YoonPRL2012}
J.~W. Yoon, G.~M. Koh, S.~H. Song, and R.~Magnusson, \enquote{Measurement and
  modeling of a complete optical absorption and scattering by coherent surface
  plasmon-polariton excitation using a silver thin-film grating,} Phys. Rev.
  Lett. \textbf{109}, 257402 (2012).

\bibitem{ZhangOE2014}
J.~Zhang, C.~Guo, K.~Liu, Z.~Zhu, W.~Ye, X.~Yuan, and S.~Qin, \enquote{Coherent
  perfect absorption and transparency in a nanostructured graphene film,} Opt.
  Express \textbf{22}, 12524--12532 (2014).

\bibitem{RaoOL2014}
S.~M. Rao, J.~J.~F. Heitz, T.~Roger, N.~Westerberg, and D.~Faccio,
  \enquote{Coherent control of light interaction with graphene,} Opt. Lett.
  \textbf{39}, 5345--5347 (2014).

\bibitem{SociNatComm2015}
T.~Roger, S.~Vezzoli, E.~Bolduc, J.~Valente, J.~J. Heitz, J.~Jeffers, C.~Soci,
  J.~Leach, C.~Couteau, N.~I. Zheludev \emph{et~al.}, \enquote{Coherent perfect
  absorption in deeply subwavelength films in the single-photon regime,} Nature
  communications \textbf{6} (2015).

\bibitem{GroteOL2013}
R.~R. Grote, J.~B. Driscoll, and J.~Richard M.~Osgood, \enquote{Integrated
  optical modulators and switches using coherent perfect loss,} Opt. Lett.
  \textbf{38}, 3001--3004 (2013).

\bibitem{BruckOE2013}
R.~Bruck and O.~L. Muskens, \enquote{Plasmonic nanoantennas as integrated
  coherent perfect absorbers on soi waveguides for modulators and all-optical
  switches,} Opt. Express \textbf{21}, 27652--27661 (2013).

\bibitem{DeLeonJSTQE2012}
N.~P. {De Leon}, M.~D. Lukin, and H.~Park, \enquote{{Quantum plasmonic
  circuits},} IEEE Journal on Selected Topics in Quantum Electronics
  \textbf{18}, 1781--1791 (2012).

\bibitem{HeeresNatNano2013}
R.~W. Heeres, L.~P. Kouwenhoven, and V.~Zwiller, \enquote{{Quantum interference
  in plasmonic circuits.}} Nature nanotechnology \textbf{8}, 719--22 (2013).

\bibitem{GroteCLEO2015}
R.~R. Grote, J.~M. Rothenberg, J.~B. Driscoll, and R.~M. Osgood, \enquote{Dpsk
  demodulation using coherent perfect absorption,} in \enquote{CLEO: Science
  and Innovations,}  (Optical Society of America, 2015), pp. STh4F--1.

\bibitem{TillmannNatPhot2013}
M.~Tillmann, B.~Daki{\'c}, R.~Heilmann, S.~Nolte, A.~Szameit, and P.~Walther,
  \enquote{Experimental boson sampling,} Nature Photonics \textbf{7}, 540--544
  (2013).

\bibitem{SpagnoloNatPhot2014}
N.~Spagnolo, C.~Vitelli, M.~Bentivegna, D.~J. Brod, A.~Crespi, F.~Flamini,
  S.~Giacomini, G.~Milani, R.~Ramponi, P.~Mataloni \emph{et~al.},
  \enquote{Experimental validation of photonic boson sampling,} Nature
  Photonics \textbf{8}, 615--620 (2014).

\bibitem{BarnettPRA1998}
S.~M. Barnett, J.~Jeffers, A.~Gatti, and R.~Loudon, \enquote{Quantum optics of
  lossy beam splitters,} Phys. Rev. A \textbf{57}, 2134--2145 (1998).

\bibitem{MojahediLPR2014}
M.~Z. Alam, J.~S. Aitchison, and M.~Mojahedi, \enquote{A marriage of
  convenience: Hybridization of surface plasmon and dielectric waveguide
  modes,} Laser and Photonics Reviews \textbf{8}, 394--408 (2014).

\bibitem{KleinScience2006}
M.~W. Klein, C.~Enkrich, M.~Wegener, and S.~Linden, \enquote{Second-harmonic
  generation from magnetic metamaterials,} Science \textbf{313}, 502--504
  (2006).

\bibitem{CaiScience2011}
W.~Cai, A.~P. Vasudev, and M.~L. Brongersma, \enquote{Electrically controlled
  nonlinear generation of light with plasmonics,} Science \textbf{333},
  1720--1723 (2011).

\bibitem{SederbergPRL2015}
S.~Sederberg and A.~Y. Elezzabi, \enquote{{Coherent Visible-Light-Generation
  Enhancement in Silicon-Based Nanoplasmonic Waveguides via Third-Harmonic
  Conversion},} Physical Review Letters \textbf{114}, 227401 (2015).

\bibitem{SegalNatPhot2015}
N.~Segal, S.~Keren-Zur, N.~Hendler, and T.~Ellenbogen, \enquote{{Controlling
  light with metamaterial-based nonlinear photonic crystals},} Nature Photonics
  \textbf{9}, 180--184 (2015).

\bibitem{KnightScience2010}
M.~W. Knight, H.~Sobhani, P.~Nordlander, and N.~J. Halas,
  \enquote{{Photodetection with active optical antennas},} Science
  \textbf{332}, 702 (2010).

\bibitem{BrongersmaNatNano2015}
M.~L. Brongersma, N.~J. Halas, and P.~Nordlander, \enquote{{Plasmon-induced hot
  carrier science and technology},} Nature Publishing Group \textbf{10}, 25--34
  (2015).

\bibitem{KonstantatosACSPhot2015}
F.~P. García~de Arquer, A.~Mihi, and G.~Konstantatos, \enquote{Large-area
  plasmonic-crystal–hot-electron-based photodetectors,} ACS Photonics
  \textbf{2}, 950--957 (2015).

\bibitem{BaldacciOE2015}
L.~Baldacci, S.~Zanotto, G.~Biasiol, L.~Sorba, and A.~Tredicucci,
  \enquote{Interferometric control of absorption in thin plasmonic
  metamaterials: general two port theory and broadband operation,} Opt. Express
  \textbf{23}, 9202--9210 (2015).

\bibitem{taylor2015enhanced}
M.~A. Taylor, M.~Waleed, A.~B. Stilgoe, H.~Rubinsztein-Dunlop, and W.~P. Bowen,
  \enquote{Enhanced optical trapping via structured scattering,} Nature
  Photonics \textbf{9}, 669 (2015).

\bibitem{Note1}
Note that symmetry and reciprocity imply $t_{11} = t_{22}$ and $t_{12} =
  t_{21}$.

\end{thebibliography}
\end{document}